\documentclass[sigconf]{acmart}

\usepackage{array} 
\usepackage{siunitx}
\usepackage{threeparttable}
\usepackage{amsmath,amsfonts}
\usepackage{enumitem}

\usepackage[super]{nth}
\usepackage{layouts}
\usepackage{float}

\usepackage{siunitx}
\usepackage{algorithmic}
\usepackage{graphicx}
\usepackage{textcomp}
\usepackage{xcolor}
\usepackage{pifont}
\usepackage{makecell}
\usepackage{flushend}
\usepackage{xfrac}
\usepackage[ruled, linesnumbered]{algorithm2e} 
\usepackage[color=pink]{todonotes}
\usepackage{color,soul}
\usepackage{multirow}
\usepackage{colortbl}
\usepackage{flushend}

\copyrightyear{2025}
\acmYear{2025}
\setcopyright{rightsretained}
\acmConference[ASPDAC '25]{30th Asia and South Pacific Design Automation
Conference}{January 20--23, 2025}{Tokyo, Japan}
\acmBooktitle{30th Asia and South Pacific Design Automation Conference
(ASPDAC '25), January 20--23, 2025, Tokyo,
Japan}\acmDOI{10.1145/3658617.3697715}
\acmISBN{979-8-4007-0635-6/25/01}

\fancypagestyle{firstpage}{%
\lhead{}
\rhead{}
\fancyhead[CO]{Accepted for publication at the 30th Asia and South Pacific Design Automation Conference
(ASPDAC '25). doi: \url{https://doi.org/10.1145/3658617.3697715}}
}

\newcommand{\red}[1]{{\color{black}#1}}


\begin{document}

\title[Design and In-training Optimization of Binary Search ADC for Flexible Classifiers]{Design and In-training Optimization of Binary \\ Search ADC for Flexible Classifiers}

\author{Paula Carolina Lozano Duarte}
\email{paula.duarte@kit.edu}
\affiliation{%
  \institution{Karlsruhe Institute of Technology}
  \country{Germany}
}

\author{Florentia Afentaki}
\email{afentaki@ceid.upatras.gr}
\affiliation{%
  \institution{University of Patras}
  \country{Greece}
 }

\author{Georgios Zervakis}
\email{zervakis@ceid.upatras.gr}
\affiliation{%
  \institution{University of Patras}
  \country{Greece}
}

\author{Mehdi B. Tahoori}
\email{mehdi.tahoori@kit.edu}
\affiliation{%
 \institution{Karlsruhe Institute of Technology}
 \country{Germany}
 }

\renewcommand{\shortauthors}{Lozano et al.}

\begin{abstract}

Flexible Electronics (FE) offer distinct advantages, including mechanical flexibility and low process temperatures, enabling extremely low-cost production. 
To address the demands of applications such as smart sensors and wearables, flexible devices must be small and operate at low supply voltages.
Additionally, target applications often require classifiers to operate directly on analog sensory input, necessitating the use of Analog to Digital Converters (ADCs) to process the sensory data.
However, ADCs present serious challenges, particularly in terms of high area and power consumption, especially when considering stringent area and energy budget.
In this work, we target common classifiers in this domain such as MLPs and SVMs and present a holistic approach to mitigate the elevated overhead of analog to digital interfacing in FE. 
First, we propose a novel design for Binary Search ADC that reduces area overhead $2\times$ compared with the state-of-the-art Binary design and up to $5.4\times$ compared with Flash ADC.
Next, we present an in-training ADC optimization in which we keep the bare-minimum representations required and simplifying ADCs by removing unnecessary components.
Our in-training optimization further reduces on average the area in terms of transistor count of the required ADCs by $5\times$ for less than 1\% accuracy loss.

\end{abstract}

\maketitle
\pagestyle{plain}

\thispagestyle{firstpage}
\section{Introduction}\label{sec:intro}

Flexible Electronics (FE) is a highly multidisciplinary research area with the potential for significant breakthroughs in developing new applications and products for ubiquitous electronics.
This field focuses on creating electronic devices that can bend, stretch, and conform to various shapes, offering advantages such as mechanical flexibility, lightweight, and low-cost production~\cite{Bonnassieux:FPE:2021:FlexiblePrintedRoadmap}.

Silicon-based electronics have improved in power consumption and integration density thanks to advanced lithography processes.
However, meeting the demands of new application areas remains challenging due to the use of bulky substrates and the need for complex and costly manufacturing equipment.
To meet the needs of emerging electronics, significant progress has been made in developing innovative materials and fabrication techniques, and FE is one of the most promising approaches in this field~\cite{Haibin:PrintedNeuromorphic}.

The aim of FE is not to replace silicon-based technologies.
In the last two decades, FE have undergone significant advancements, resulting in the development of mature, low-cost, thin, flexible, and conformable devices.
These devices encompass a wide array of components, including sensors, memories, batteries, light-emitting diodes (LEDs), energy harvesters, near-field communication (NFC)/radio frequency identification (RFID) modules, or antennas~\cite{ARM:2021:PlasticARM, chang2017circuits, cui2016printed}.
These components constitute the foundational building blocks for constructing various types of smart integrated electronic devices.
Applications span across multiple domains, such as healthcare (e.g., wearable health monitors and smart bandages), forensics (e.g., flexible fingerprint sensors), packaging (e.g., smart labels and tags for inventory management), and other promising fields like consumer electronics and environmental monitoring~\cite{FlexibleElectronics}.

\begin{figure}[!t]
\centering
\includegraphics[width=0.5\textwidth]{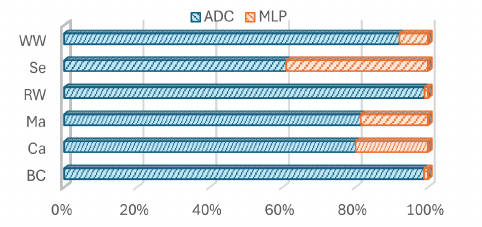}
\vspace{-6ex}
\caption{Transistor Count Evaluation of the printed classification system.}
\vspace{-5ex}
\label{fig:motivation}
\end{figure}

Sensor processing applications are among the most appropriate applications of FE.
The crucial component for sensor processing is the analog-to-digital converter (ADC). 
But implementing ADCs in FE faces significant challenges, particularly in terms of high area and power consumption since they require numerous components to perform the conversion process.
\figurename~\ref{fig:motivation} presents the ratio of the ADC in the classification system that uses the MLPs proposed in~\cite{Afentaki:DATE2024}.
As can be observed the ADCs are the dominant source of area and power in a digital processing system, making them a primary bottleneck in achieving efficient, compact, and low-power FE devices.

In this work first, we incorporate the idea of the Binary Search into the flash-type ADC design.
This approach enables eliminating the costly digital encoder from the flash ADC, saving considerable amount of power and area.
Secondly, we propose our Binary Search ADC design contains two main parts: i. Comparator and ii. Control unit.
The comparators are fundamentally required by the flash-type ADC and the control unit is required to activate the respective comparator based on the Binary Search algorithm.
Third, we propose a high-level in-training optimization which taking into account the target application i.e., MLP classification system, highly customized pruned Binary ADCs are obtained.

The rest of the paper is structured as follows.
Section~\ref{sec:background} provides an overview of FE, its advantages and limitations, and a discussion of existing flexible ADCs available in the literature.
In Section~\ref{sec:framework}, we outline our proposed design for the Binary Search ADC, analyzing each essential component. 
Also, the high-level ADC pruning optimization methodology is described for obtaining highly customized pruned ADCs.
Section~\ref{sec:results} presents a discussion of the simulation results and evaluations is obtained.
And finally, Section~\ref{sec:conclusion} consolidates the study's findings and draws conclusive insights.

\section{Background}\label{sec:background}
\subsection{Flexible Electronics}\label{subsec:PE_FE}

\textit{Flexible electronics} (FE) include electronic devices or systems capable of bending, stretching, or deforming without impairing their functionality.
Typically fabricated on flexible substrates like plastic, rubber, or thin metal foils, these devices adapt to diverse shapes and surfaces.
Promising applications for flexible electronics include lightweight, portable, and conformable electronic systems, such as wearable health monitors or bendable sensors~\cite{Gao:FlexibleWearableSensing}.
Traditional semiconductor manufacturing techniques can also be used to create flexible electronics by using flexible substrates instead of rigid ones.
Pragmatic Semiconductor, for instance, utilizes a streamlined version of silicon lithography for flexible electronics, significantly reducing costs by eliminating several complex steps typical in traditional silicon-based manufacturing, such as ion implantation and high-temperature annealing.
Instead, it employs automated, self-contained production lines that ensure high-volume, cost-effective fabrication of flexible integrated circuits (FlexICs)~\cite{FlexICs}.

In this work we consider the subtractive flexible technology of PragmatIC FlexICs technology, which employs photolithographic techniques to achieve critical Thin-Film Transistor (TFT) dimensions of 600nm.
This platform utilizes a thin, high-k gate dielectric, allowing operation at standard Integrated Circuit (IC) input voltages.
Additionally, it integrates a dedicated high-value resistor component and offers four metal layers, 
all manufactured on a polyimide substrate~\cite{FlexICs}.
This technology provides \textit{only} N-type transistors, imposing constraints on circuit design.

\subsection{Analog to Digital Converter}\label{subsec:ADC}

Analog-to-Digital Converters (ADCs) are essential components in electronics, as they serve as the interface between analog and digital domain.
There are various types of  ADC architectures~\cite{Sedra:microelectronics}, each with its advantages and disadvantages, tailored to specific application requirements.

One common type of ADC is the $N$-bit Flash ADC which utilizes $2^{N}\!-\!1$ comparators to compare the input analog voltage with multiple reference voltages simultaneously.
Each comparator produces a digital output indicating whether the input voltage is greater or less than the corresponding reference voltage.
The digital outputs from all comparators are then encoded to generate the final digital output.
Flash ADCs offer high speed for the conversion of input signals, making them suitable for applications requiring rapid sampling rates, but they consume large power and area.
As a result, implementing high-resolution Flash ADC is costly~\cite{JAMSHIDIROUDBARI:2010:FlashFlexible}. 

Binary Search ADCs are commonly used in applications requiring low power and easy implementation.
They digitize the input voltage through a series of cascaded stages, each performing a binary search.
As each stage refines the digital output with increasing precision, this leads to a high-resolution conversion.
This architecture is particularly advantageous for its low area and straightforward implementation compared to other ADC types~\cite{Bhai:CMI:2016:desingbinADCNcomp}.

Another type of ADC is the Successive Approximation Register (SAR) ADC.
SAR ADCs use a binary search algorithm to approximate the input analog signal.
The SAR ADC compares the input signal with a reference voltage and iteratively refines the approximation until the digital output matches the input signal’s value.
SAR ADCs offer a balance of low power consumption and medium resolution, making them suitable for various applications, including precision measurement and data acquisition systems~\cite{Alkhalil:BioCAS:2022:FlexibleSAR}.

Lastly, Delta-Sigma (\(\Delta\Sigma\)) ADCs are another type of ADC architecture.
It uses oversampling and noise shaping techniques to achieve high-resolution digital output.
These ADCs convert the input signal into a stream of pulses, where the pulse density represents the signal amplitude.
By oversampling the input signal at a high frequency and filtering out the quantization noise, Delta-Sigma ADCs can achieve high-resolution digital output with low distortion.
Delta-Sigma ADCs are commonly used in audio applications, such as digital audio recording and playback systems~\cite{Garripoli:2017:DeltaSigmaADC}.

\begin{table}[]

\caption{Comparison between available ADC. Scale 1-4 where 1 represents the lowest in area, power or speed}\vspace{-2ex}\label{fig:comADC} 
\scalebox{0.8}{
    \begin{tabular}{|c|c|c|c|c|}
\hline
\rowcolor[HTML]{C0C0C0} 
{\color[HTML]{000000} Architecture} & {\color[HTML]{000000} Area} & {\color[HTML]{000000} Power} & {\color[HTML]{000000} Speed} & {\color[HTML]{000000} Application}  \\ \hline
Flash                               & 4                           & 4                            & 4                            & High   speed and resolution         \\ \hline
SAR                                 & 3                           & 3                            & 3                            & Low   power and medium resolution   \\ \hline
Sigma-Delta                         & 2                           & 1                            & 1                            & Low   power and high resolution     \\ \hline
Binary                              & 1                           & 2                            & 2                            & Low   power and easy implementation \\ \hline
\end{tabular}}
\vspace{-2ex}
\end{table}

In summary, Flash ADCs provide high speed but are costly due to their large power and area consumption.
SAR ADCs offer a balance of high resolution and low power consumption, making them ideal for precision measurement and data acquisition systems.
Delta-Sigma ADCs excel in high-resolution and low-distortion applications.
Binary Search ADCs are advantageous for their low area and straightforward implementation, providing a high-resolution conversion with low power consumption.
The choice of ADC architecture depends on the specific requirements of the application, balancing factors such as speed, resolution, power consumption, and cost.
In this work, we focus on exploring innovative approaches to achieve low power consumption and optimize area utilization in the context of Binary Search ADCs compared with Flash ADC.
In Table~\ref{fig:comADC} we summarize the advantages, disadvantages, and application of the described ADCs.

Based on the observation that many levels of comparison are unused, it is possible to create partial/pruned ADCs that reduce quantization levels without compromising accuracy. To achieve this, we first develop a training algorithm that minimizes these levels, followed by the design of the customized target ADC.

Research on flexible ADCs is limited.
For example,~\cite{JAMSHIDIROUDBARI:2010:FlashFlexible} presents a Flash ADC,~\cite{Alkhalil:BioCAS:2022:FlexibleSAR} proposes a SAR ADC, and~\cite{Garripoli:2017:DeltaSigmaADC} presents a Sigma Delta ADC.
To the best of our knowledge, this is the first work that proposes the design of a binary-search-like ADC for FE.
As for the exploration of partial designs, the research is more limited~\cite{Armeniakos:DATE24}, and it explores only Flash ADC.

As the baseline for the binary search ADC, we consider the design proposed by~\cite{Bhai:CMI:2016:desingbinADCNcomp}.
As illustrated in Figure~\ref{fig:ReferenceBinADC} part a), the 3 Bit Binary Search ADC comprises 3 comparators, 2 NOT, 4 AND gates and 6 transistors.
This ADC is designed for CMOS technology and utilizes both P and N type transistors.
For a fair comparison between designs, we adapted it to use only N-type transistors.
\section{Optimizing Binary Search ADC Architecture}\label{sec:framework}

\subsection{Binary Search ADC for area constrained applications}\label{subsec:proposed_design}

\begin{figure}[!t]
\centering
\includegraphics[width=\columnwidth]{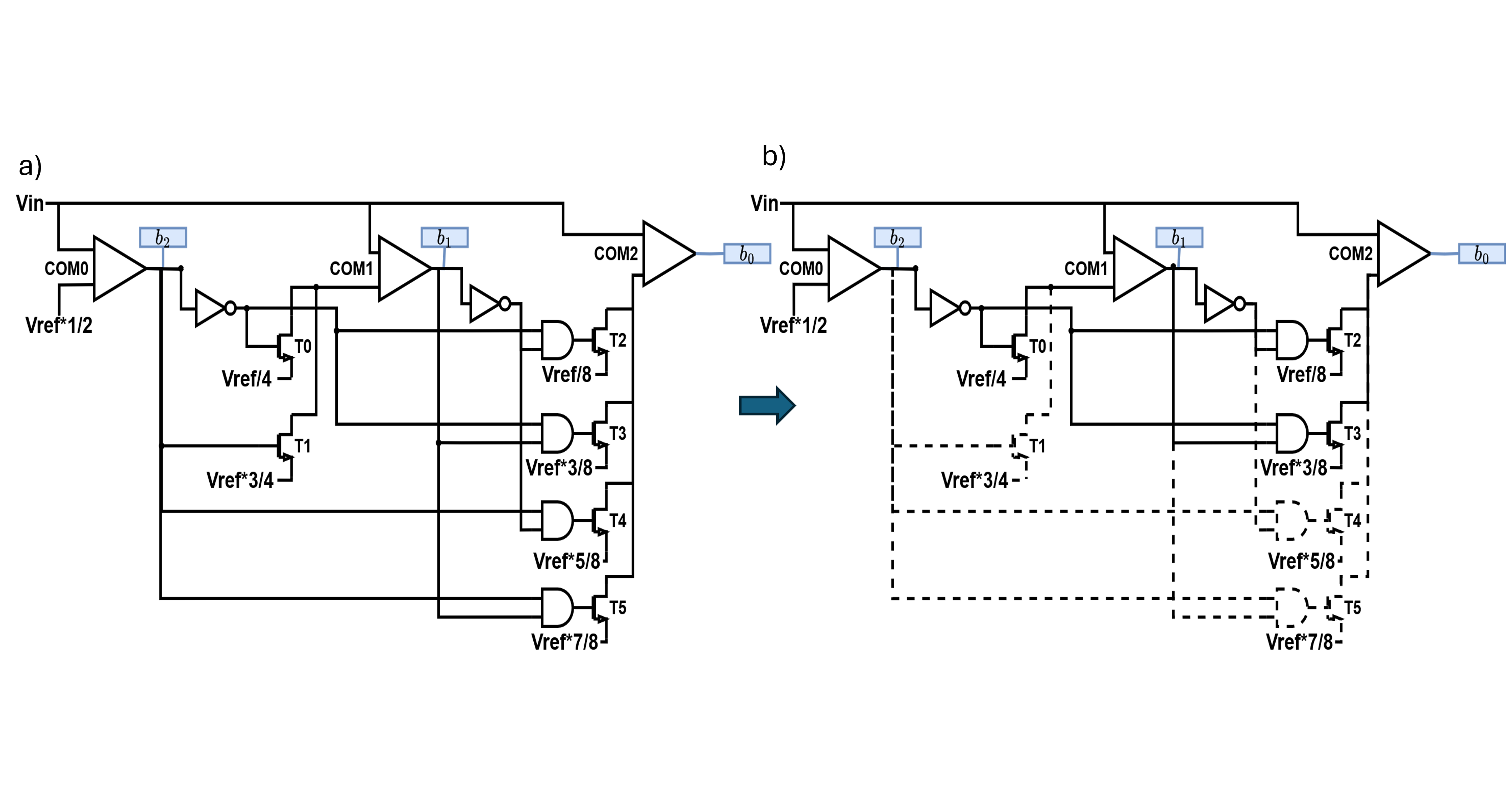}
\vspace{-10ex}
\caption{Schematic of: a) conventional 3-bit Binary ADC and b) an example of an equivalent bespoke partial/pruned ADC}
\vspace{-3ex}
\label{fig:afentaki}
\label{fig:ReferenceBinADC}
\end{figure}

\begin{figure*}
    \centering
    \includegraphics[width=1\linewidth]{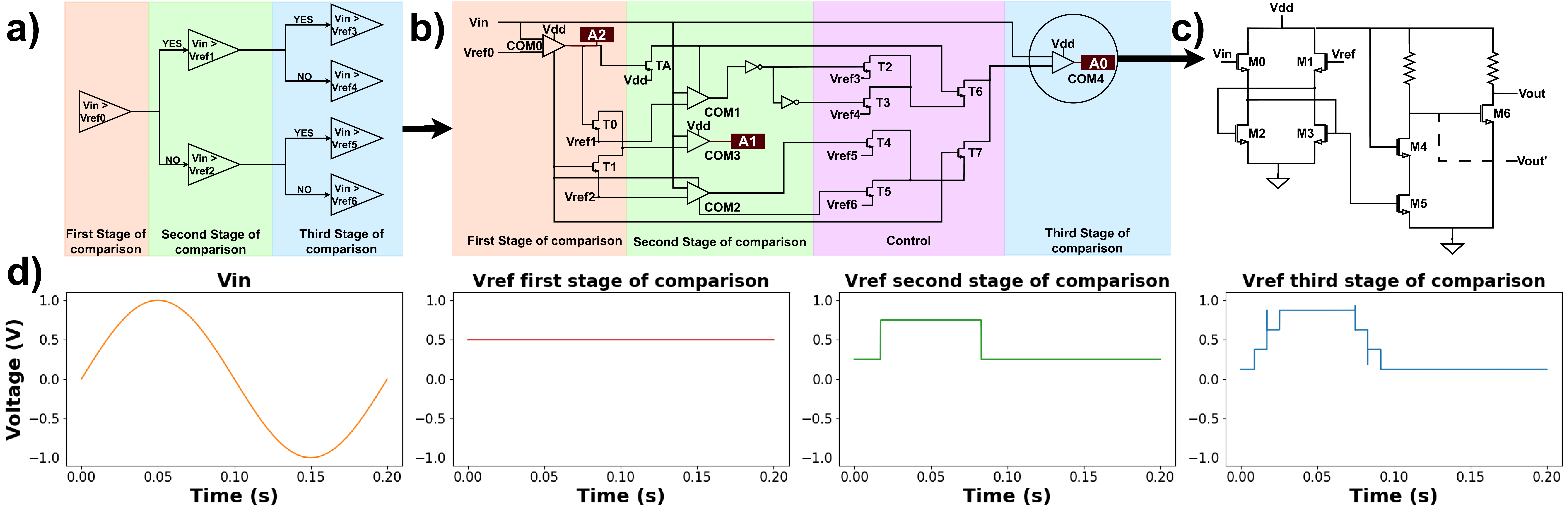}
    \caption{Proposed Design. a) Block diagram of a Binary Search ADC. b) Schematic of the proposed design. c) Schematic of the comparator. d) $V_{in}$, $V_{ref}$ in the first stage of comparison, $V_{ref}$ in the second stage of comparison, $V_{ref}$ in the third stage of comparison}
    \label{fig:BinADCProposed} \vspace{-2ex}
    \Description[Proposed design for the Binary Search ADC]
    \vspace{}
\end{figure*}
\noindent

A $N$-bits Binary Search ADC has $N$ stages of comparison between the input signal and a reference voltage that is determined by the previous stage, see Figure~\ref{fig:BinADCProposed} a).
The idea is to reduce power consumption by only using half of the comparators compared with a Flash ADC. 
Given the resource-constrained nature of FE circuits, our main objective is to propose an FE-specific ADC design while optimizing the number of required components, and as illustrated in Figure~\ref{fig:BinADCProposed} b), our proposed design for a 3-bit Binary Search ADC comprises 5 comparators, 2 NOT, and 9 transistors.

The number of comparators, inverters and transistors increase with respect to the bit-width of the ADC.
However, although the number of comparators needed for specific bit-widths could be formulated; there are no straightforward formulas to calculate the numbers for inverters and transistors.
As we explain below with the example of the 3-bit ADC, for higher {$V_{ref}$}, it may be needed to add extra transistors, like TA, or double inversions, like the ones after COM1. 

The operational principle of circuit is as follows: 

\begin{itemize}
    {\item \textbf{First stage of comparison:} with this first comparison we get our Most Significant Bit (MSB), A2.
    The input is compared with $V_{ref0}$ (\textit{$V_{ref}/2$}).
    If the input is greater, this 1 is used as the supply for the comparator COM1, working as an enable/disable; otherwise, a 1 is sent to power the comparator COM2.
    Consequently, only one of the comparators COM1 or COM2 receives power and is enabled.
    The purpose of the transistors T0 and T1 is to create a new reference voltage for COM3.
    If the output of COM0 is 1 ($V_{in}$ > $V_{ref0}$), T0 is ON and the new reference voltage is $V_{ref1}$, otherwise when the output in COM0 is 0 the gate in T1 goes to ON and provide $V_{ref2}$ as reference voltage for COM3.
    In this way, the reference voltage for COM3 is either $V_{ref1}$ or $V_{ref2}$, depending if $V_{in}$ is higher than $V_{ref0}$.}
    
    \item \textbf{Second stage of comparison:} the output of COM3 is A1. the comparison involves comparing against $V_{ref1}$ (\textit{$V_{ref}\cdot \sfrac{3}{4}$}) in COM1 or $V_{ref2}$ (\textit{$V_{ref}/4$}) in COM2, giving as a result an enable and disable for the next stage.
    In COM3 the comparison is between $V_{in}$ and $V_{ref1}$ or $V_{ref2}$, as explained previously.
    Furthermore, before the comparison of the higher voltages ($V_{ref3}$ and $V_{ref4}$) we need to add TA to amplify the output of COM0.
    TA goes ON when A2 is 1, allowing $V_{dd}$ to go through the drain which is connected to the $V_{dd}$ of the next comparator.

    \item \textbf{Control block:} It is made with $2^{N}\!-\!2$ transistors, 6 for a 3 bit ADC, and it selects the reference voltage, either $V_{ref3}$, $V_{ref4}$, $V_{ref5}$ or $V_{ref6}$.
    To accomplish this, T2 and T3 provide a reference signal choosing between $V_{ref3}$ (\textit{$V_{ref}\cdot \sfrac{7}{8}$}) and $V_{ref4}$ (\textit{$V_{ref}\cdot \sfrac{5}{8}$}), while T4 and T5 is either $V_{ref5}$ (\textit{$V_{ref}\cdot \sfrac{3}{8}$}) or $V_{ref6}$ (\textit{$V_{ref}/8$}).
    T6 and T7 aid in selecting the appropriate signal based on whether COM0 > $V_{ref0}$ or not.

    \item \textbf{Third stage of comparison:} COM4 is comparing the input signal with the 4 possibles levels of the reference voltage ($V_{ref3}$, $V_{ref4}$, $V_{ref5}$ or $V_{ref6}$). Its output is A0, our Least Significant Bit (LSB).
    In Figure~\ref{fig:BinADCProposed} part d), we show the reference voltage of each comparison stage.
    
\end{itemize}

\textit{Comparator}\label{subsec:comparator}: The schematic of the proposed comparator is depicted in Figure~\ref{fig:BinADCProposed} c).
It features two outputs: one representing the regular output, $V_{out}$, while $V_{out'}$ presents its inverse.
This design choice is deliberate, as it ensures a more streamlined approach compared to employing an inverter post-comparison.
By incorporating inversion directly within the comparator structure, we optimize the sequential flow of operations.
Our comparator utilizes of 7 transistors, 2 resistors and it works as follows: transistors M0 and M1 enable the comparison when having $V_{dd}$.
It is specially important that both transistors are the same size to ensure that $V_{in}$ and $V_{ref}$ are treated equally.
Similarly, M2 and M3 should also be matched.
Transistors M4, M5, M6 and the two resistors force the output of the comparison to be 0 or 1.

One of the main challenges we need to address is designing a comparator using only N-type transistors.
Additionally, it is crucial to determine the appropriate sizes for all transistors, as the width-to-length $(\sfrac{W}{L})$ ratios between M0 and M1, and M2 and M3, influence the comparators behavior, primarily based on the inputs.
Generally, a higher $(\sfrac{W}{L})$ for M2 and M3 enhances the comparators outputs.

Comparing the baseline~\ref{fig:ReferenceBinADC} with our design~\ref{fig:BinADCProposed}, their switching network uses 2 inverters and 4 ANDs, while ours uses 6 transistors, called the control block. Although they need 3 comparators, and we require 5, our design reduces area overhead by half while maintaining the same power, as shown in the results.

The advantages of our design include the exclusive use of N-type transistors and resistors, as required by FE technologies, with a control block made entirely of transistors, minimizing power and area. Additionally, the hierarchical design allows easy scalability for higher precision ADCs. The main drawback is lower speed compared to Flash ADCs, but speed is not critical for the target applications, which operate from a few Hz to a few kHz~\cite{Bleier:ISCA:2020:printedmicro}.

\subsection{High-level Optimization Pruning Binary Search ADCs}\label{subsec:pruning}

In this work, we move one step further, offering a second optimization option by transitioning from generic to bespoke solutions. 
FE allows designing fully customized partial/pruned Binary (ADCs). 
This allows us to create bespoke ADCs tailored specifically for each sensor input, optimized according to the unique requirements of the dataset and model. 
Utilizing FE technology enables us to achieve these bespoke designs efficiently and cost-effectively, providing significant flexibility and precision in sensor data acquisition.
The above has motivated us to explore more fine-grained solutions, examining which combinations of the $2^N$ ADCs representation, where $N$ is the ADC resolution, most accurately and efficiently describe the input. 
This is achieved through ADC pruning, where only the necessary representations for each sensor input are retained, resulting in the removal of unused input representations and their corresponding circuitry, i.e., sub-trees of binary search ADC and their corresponding comparators.

The pruning procedure reduces the number of the ADCs sub-components by removing the unnecessary representation levels. 
In a conventional binary $N$-bit ADC, the analog input voltage ($V_{in}$) is compared among $N$ stage of comparison with a reference voltage value($V_{ref}$) which is selected by the control block.

During the pruning procedure, a subset of quantization levels is selected while the remaining levels are pruned.
By removing unnecessary levels, the number of sub-components inside each level in the final pruned binary ADC is equal to the number of remaining levels.
This reduction in the number of sub-components leads to a decrease in the complexity of the ADC circuitry, resulting in area and power savings.
More specific, let $a$ be the quantization level that will be removed.
During the pruning, the transistor that holds the corresponding $V_{ref}$ of the $a$ value is removed.
It can be observed that if the pruning occurs in the control block of the third-stage of comparison an extra AND gate is also removed.
Further, if all the values of the same level are pruned then a comparator is also pruned.
If the pruning occurs in $V_{ref}/2$ the first-order comparator will be removed and subsequently this will lead to pruning half the tree.
The above consist the set of design rules that are used in order to design and sculpture a high-level area model for the partial binary ADCs.
The latter will be further discussed later in the paper.

The representation pruning, leads to accuracy degradation of the model, especially if course-grain techniques would have been applied that may cause in removing important input's representation, vital for the characterization of the input's distribution.
Ideally, the optimal pruned ADC would be the one that follows the input's distribution since the optimal ADC will have the minimum number of representations to retain maximum classification accuracy.
Thus, after the pruning exploration a quantization-aware training is applied, in order to evaluate the solutions that have the best classification accuracy and at the same time the minimum ADC area.
This multi-objective problem is formed as the maximization of the classification accuracy and the minimization of the ADC area i.e.,  $min\{1-accuracy, ADC_{area}\}$.

This multi-objective exploration problem is explored using a genetic algorithm.
We utilize the Non-dominated Sorting Genetic Algorithm II (NSGA-II) to solve this problem because to its simplicity, low computational complexity, and improved convergence ~\cite{Deb:NSGA:2002}.
NSGA-II incorporates elitism through the fast-non-dominated-sort method, which improves the convergence of the exploration.
The genetic algorithm (GA) used in the proposed ADC-aware methodology encodes the parameters of the ADC pruning as well as the decimal point position of the coefficients in the classifier's quantization-aware training (QAT).
The encoding process represents the possible solutions or individuals as binary strings, with each bit representing a quantization level, indicating whether it is kept or pruned.
The GA works on a population of binary-encoded individuals, using genetic operators like selection, crossover, and mutation to evolve the population over time.
The fitness of each individual is assessed using the performance of the corresponding pruned ADC in terms of accuracy and resource utilisation.
By iteratively applying genetic operators and evaluating fitness, the GA explores the search space to find an optimal or near-optimal ADC configuration that balances accuracy and resource efficiency.

In order to capture the area of the ADC during the genetic exploration the quantification of ADC area reduction is needed.
However, using a LUT-technique, that creates a dictionary between the mask parameters and the partial ADC area would require designing all the ADC pruning combinations, which is not efficient in terms of time as the ADC resolution increases.
The area model in the proposed ADC-pruning methodology uses a high-level area model, which estimates the area of the pruned bespoke ADCs.
The area model uses the mask parameters that form the partial ADC and calculates the ADC area based on the aforementioned design rules of the binary ADC.

By taking these into account, our model calculates the area reduction accomplished by the pruning method, allowing for efficient exploration of the ADC design space in the evolutionary algorithm.

\section{Simulation results and Evaluation}\label{sec:results}

\subsection{Experimental Setup}\label{sec:experimental_setup}

\begin{table}[]
\caption{Design properties Binary ADC}\label{fig:sizesDevices}\vspace{-2ex}
\scalebox{0.88}{\begin{tabular}{ccc}
\hline
\rowcolor[HTML]{C0C0C0} 
\multicolumn{2}{|c|}{\cellcolor[HTML]{C0C0C0}{\color[HTML]{000000} Component}}                                                                   & \multicolumn{1}{c|}{\cellcolor[HTML]{C0C0C0}{\color[HTML]{000000} Size}} \\ \hline
\rowcolor[HTML]{E7E7E7} 
\multicolumn{1}{|c|}{\cellcolor[HTML]{FFFFFF}}                             & \multicolumn{1}{c|}{\cellcolor[HTML]{E7E7E7}M0, M1}                 & \multicolumn{1}{c|}{\cellcolor[HTML]{E7E7E7}W= 2$\mu$m L= 600nm}             \\ \cline{2-3} 
\rowcolor[HTML]{FFFFFF} 
\multicolumn{1}{|c|}{\cellcolor[HTML]{FFFFFF}}                             & \multicolumn{1}{c|}{\cellcolor[HTML]{FFFFFF}M2 ,M3}                 & \multicolumn{1}{c|}{\cellcolor[HTML]{FFFFFF}W= 4$\mu$m L= 600nm}             \\ \cline{2-3} 
\rowcolor[HTML]{E7E7E7} 
\multicolumn{1}{|c|}{\cellcolor[HTML]{FFFFFF}}                             & \multicolumn{1}{c|}{\cellcolor[HTML]{E7E7E7}M4, M5, M6}             & \multicolumn{1}{c|}{\cellcolor[HTML]{E7E7E7}W= 8$\mu$m L= 600nm}             \\ \cline{2-3} 
\rowcolor[HTML]{FFFFFF} 
\multicolumn{1}{|c|}{\multirow{-4}{*}{\cellcolor[HTML]{FFFFFF}COM0, COM2}} & \multicolumn{1}{c|}{\cellcolor[HTML]{FFFFFF}R = 70,057MΩ}           & \multicolumn{1}{c|}{\cellcolor[HTML]{FFFFFF}W= 1,4$\mu$m L= 510$\mu$m}           \\ \hline
\rowcolor[HTML]{E7E7E7} 
\multicolumn{1}{|c|}{\cellcolor[HTML]{FFFFFF}}                             & \multicolumn{1}{c|}{\cellcolor[HTML]{E7E7E7}M0, M1, M2, M3, M4, M5} & \multicolumn{1}{c|}{\cellcolor[HTML]{E7E7E7}W= 2$\mu$m L= 600nm}             \\ \cline{2-3} 
\rowcolor[HTML]{FFFFFF} 
\multicolumn{1}{|c|}{\multirow{-2}{*}{\cellcolor[HTML]{FFFFFF}COM1*}}      & \multicolumn{1}{c|}{\cellcolor[HTML]{FFFFFF}R = 70,057MΩ}           & \multicolumn{1}{c|}{\cellcolor[HTML]{FFFFFF}W= 1,4$\mu$m L= 86,2$\mu$m}          \\ \hline
\rowcolor[HTML]{E7E7E7} 
\multicolumn{1}{|c|}{\cellcolor[HTML]{FFFFFF}}                             & \multicolumn{1}{c|}{\cellcolor[HTML]{E7E7E7}M0, M1}                 & \multicolumn{1}{c|}{\cellcolor[HTML]{E7E7E7}W= 5$\mu$m L= 600nm}             \\ \cline{2-3} 
\rowcolor[HTML]{FFFFFF} 
\multicolumn{1}{|c|}{\cellcolor[HTML]{FFFFFF}}                             & \multicolumn{1}{c|}{\cellcolor[HTML]{FFFFFF}M2 ,M3}                 & \multicolumn{1}{c|}{\cellcolor[HTML]{FFFFFF}W= 2$\mu$m L= 600nm}             \\ \cline{2-3} 
\rowcolor[HTML]{E7E7E7} 
\multicolumn{1}{|c|}{\cellcolor[HTML]{FFFFFF}}                             & \multicolumn{1}{c|}{\cellcolor[HTML]{E7E7E7}M4, M5, M6}             & \multicolumn{1}{c|}{\cellcolor[HTML]{E7E7E7}W= 2$\mu$m L= 800nm}             \\ \cline{2-3} 
\rowcolor[HTML]{FFFFFF} 
\multicolumn{1}{|c|}{\multirow{-4}{*}{\cellcolor[HTML]{FFFFFF}COM3, COM4}} & \multicolumn{1}{c|}{\cellcolor[HTML]{FFFFFF}R = 70,057MΩ}           & \multicolumn{1}{c|}{\cellcolor[HTML]{FFFFFF}W= 1,4$\mu$m L= 510$\mu$m}           \\ \hline
\rowcolor[HTML]{E7E7E7} 
\multicolumn{1}{|c|}{\cellcolor[HTML]{FFFFFF}}                             & \multicolumn{1}{c|}{\cellcolor[HTML]{E7E7E7}M}                      & \multicolumn{1}{c|}{\cellcolor[HTML]{E7E7E7}W= 3$\mu$m L= 600nm}             \\ \cline{2-3} 
\rowcolor[HTML]{FFFFFF} 
\multicolumn{1}{|c|}{\multirow{-2}{*}{\cellcolor[HTML]{FFFFFF}INV}}        & \multicolumn{1}{c|}{\cellcolor[HTML]{FFFFFF}R = 26,4MΩ}             & \multicolumn{1}{c|}{\cellcolor[HTML]{FFFFFF}W= 1,4$\mu$m   L=194,6$\mu$m}        \\ \hline
\rowcolor[HTML]{E7E7E7} 
\multicolumn{2}{|c|}{\cellcolor[HTML]{FFFFFF}T0,T1,T2,T3,T4,T5,T6,T7, TA}                                                                        & \multicolumn{1}{c|}{\cellcolor[HTML]{E7E7E7}W= 2$\mu$m L= 600nm}             \\ \hline
\multicolumn{3}{l}{*COM1 only has Vout’ as output and the second R and M6 are removed}                                                                                                                   
\end{tabular}
}
\vspace{-3ex}
\end{table}

All blocks of the ADC are simulated using the Cadence Spectre simulator. 
For the subtractive flexible technology, we use PragmatIC FlexICs PDK second generation Helvellyn 2.1.0~\cite{FlexICs}, and simulations are conducted with $V_{dd}$ of 1V, $V_{ref}$ of 1V and a temperature of 27ºC.
Our $V_{in}$ is a sine of 1V with a frequency of 5Hz.

The dimensions of the components for each device of our Binary ADC are provided in Table~\ref{fig:sizesDevices}.
For the Flash ADC, we use in the resistor ladder resistances of $1.17M\Omega$ and comparators with transistors of size $W = 5\mu m$ and $L = 600nm$ with resistors of $256.8M\Omega$.

For the Synthesis, we developed a standard-cell library with Pragmatic PDK by designing the standard cells and making the appropriate measurements.
The MLPs are synthesized using Synopsys Design Compiler S-2021.06 and mapped to the Pragmatic. 
library~\cite{Bleier:ISCA:2020:printedmicro}, while VCS T-2022.06 and PrimeTime T-2022.03 are used for simulation and power analysis.
The coefficients of the MLP are 8-bit fixed-point values.
Both the quantization parameters and the topology of the MLPs align with previous works in~\cite{Argyris:DATE2023, Afentaki:DATE2024}, ensuring a fair comparison.
The accuracy is reported on the test dataset, and all designs are synthesized at a relaxed clock period of $5\si{\kilo\hertz}$.
Such delay values align with typical FE  performance~\cite{cadilha2017digital}.
The architecture of the MLPs is the same as the authors have reported in~\cite{Mubarik:MICRO2020} and~\cite{Afentaki:ICCAD23}.
The inputs are normalized to $[0, 1]$ as in~\cite{Mubarik:MICRO2020, Afentaki:ICCAD23, Argyris:DATE2023} and are randomly stratified split into $70\%/30\%$ train/test sets, ensuring a balanced distribution of each target class in each of these sets.
Mutation and crossover operators of the GA are set to $0.2\%$ \text{and} $0.7\%$ respectively.
The open-source Pymoo framework is used for the NSGA-II implementation~\cite{Pymoo}.
Our exact baseline are the bespoke power-of-2 quantized MLP circuits, designed following the approach outlined in~\cite{Argyris:DATE2023}, using $8$-bit fixed point power-of-2 values weights.

\subsection{Full Binary ADCs Evaluation}\label{sec:comparison}

In Table~\ref{fig:FlashArea} we show the area for the Flash ADC distinguishing between the area that comes from the resistor ladder and comparators and the area of the encoder.
The principle power overhead comes from the encoder.
This finding is meaningful because with the Binary Search ADC we do not need an encoder, since the output is already digital.

\begin{table}[]
\caption{Flash ADC detail area}\label{fig:FlashArea}
\vspace{-2ex}
\scalebox{0.89}{\begin{tabular}{|
>{\columncolor[HTML]{FFFFFF}}c c|c|c|l}
\cline{1-4}
\multicolumn{2}{|c|}{\cellcolor[HTML]{C0C0C0}{\color[HTML]{000000} 3   Bits Flash ADC}}                    & \cellcolor[HTML]{C0C0C0}{\color[HTML]{000000} Resistor ladder + comparator} & \cellcolor[HTML]{C0C0C0}{\color[HTML]{000000} Encoder 7 to 3} &  \\ \cline{1-4}
\multicolumn{1}{|c|}{\cellcolor[HTML]{FFFFFF}}                        & \cellcolor[HTML]{E7E7E7}Area($\mu$m²)  & \cellcolor[HTML]{E7E7E7}85745                                               & \cellcolor[HTML]{E7E7E7}9321                                  &  \\ \cline{2-4}
\multicolumn{1}{|c|}{\multirow{-2}{*}{\cellcolor[HTML]{FFFFFF}Area}}  & \cellcolor[HTML]{FFFFFF}Overall    & \cellcolor[HTML]{FFFFFF}90,2\%                                              & \cellcolor[HTML]{FFFFFF}9,8\%                                 &  \\ \cline{1-4}
\multicolumn{1}{|c|}{\cellcolor[HTML]{FFFFFF}}                        & \cellcolor[HTML]{E7E7E7}Power (nW) & \cellcolor[HTML]{E7E7E7}462,2                                               & \cellcolor[HTML]{E7E7E7}531                                   &  \\ \cline{2-4}
\multicolumn{1}{|c|}{\multirow{-2}{*}{\cellcolor[HTML]{FFFFFF}Power}} & \cellcolor[HTML]{FFFFFF}Overall    & \cellcolor[HTML]{FFFFFF}46,54\%                                             & \cellcolor[HTML]{FFFFFF}53,46\%                               &  \\ \cline{1-4}
\end{tabular}}
\vspace{-3ex}
\end{table}

\begin{table}[]
\caption{Comparison between full ADCs}
\vspace{-2ex} \label{fig:ComFullPrintedADC}
\scalebox{0.83}{\begin{tabular}{|cc|c|c|c|}
\hline
\rowcolor[HTML]{C0C0C0} 
\multicolumn{2}{|c|}{\cellcolor[HTML]{C0C0C0}{\color[HTML]{000000} Architecture}}                                                                                                      & {\color[HTML]{000000} Accuracy   (bits)} & {\color[HTML]{000000} Area($\mu$m²)} & {\color[HTML]{000000} Power   (nW)} \\ \hline
\rowcolor[HTML]{E7E7E7} 
\multicolumn{2}{|c|}{\cellcolor[HTML]{FFFFFF}{\color[HTML]{000000} }}                                                                                                                  & {\color[HTML]{000000} 3}                 & {\color[HTML]{000000} 95066}     & {\color[HTML]{000000} 993,2}        \\ \cline{3-5} 
\rowcolor[HTML]{FFFFFF} 
\multicolumn{2}{|c|}{\multirow{-2}{*}{\cellcolor[HTML]{FFFFFF}{\color[HTML]{000000} Flash}}}                                                                                           & {\color[HTML]{000000} 4}                 & {\color[HTML]{000000} 212635}    & {\color[HTML]{000000} 2684}         \\ \hline
\rowcolor[HTML]{E7E7E7} 
\multicolumn{1}{|c|}{\cellcolor[HTML]{FFFFFF}{\color[HTML]{000000} }}                                  & \cellcolor[HTML]{FFFFFF}{\color[HTML]{000000} }                               & {\color[HTML]{000000} 3}                 & {\color[HTML]{000000} 35722}     & {\color[HTML]{000000} 365,1}        \\ \cline{3-5} 
\rowcolor[HTML]{FFFFFF} 
\multicolumn{1}{|c|}{\cellcolor[HTML]{FFFFFF}{\color[HTML]{000000} }}                                  & \multirow{-2}{*}{\cellcolor[HTML]{FFFFFF}{\color[HTML]{000000} Baseline}}     & {\color[HTML]{000000} 4}                 & {\color[HTML]{000000} 86556}     & {\color[HTML]{000000} 829,5}        \\ \cline{2-5} 
\rowcolor[HTML]{E7E7E7} 
\multicolumn{1}{|c|}{\cellcolor[HTML]{FFFFFF}{\color[HTML]{000000} }}                                  & \cellcolor[HTML]{FFFFFF}{\color[HTML]{000000} }                                        & {\color[HTML]{000000} \textbf{3}}        & {\color[HTML]{000000} \textbf{17679}} & {\color[HTML]{000000} \textbf{360}}   \\ \cline{3-5} 
\rowcolor[HTML]{FFFFFF} 
\multicolumn{1}{|c|}{\multirow{-4}{*}{\cellcolor[HTML]{FFFFFF}{\color[HTML]{000000} Binary   Search}}} & \multirow{-2}{*}{\cellcolor[HTML]{FFFFFF}{\color[HTML]{000000} \textbf{New   design}}} & {\color[HTML]{000000} \textbf{4}}        & {\color[HTML]{000000} \textbf{50027}} & {\color[HTML]{000000} \textbf{541,8}} \\ \hline
\end{tabular}}
\vspace{-3ex}    
\end{table}

In Table~\ref{fig:ComFullPrintedADC}, we compare three ADC designs: Binary ADC baseline design presented in Figure~\ref{fig:ReferenceBinADC} a), our design, and a Flash ADC.
For 3-bit, our design reduces area by $2\times$ compared to the baseline and $5.4\times$ compared to Flash.
For 4-bit, area is reduced by $1.7\times$ and $4.3\times$, respectively.
Power consumption is slightly lower than the Binary baseline and $2.8\times$ lower than Flash for 3-bit, while for 4-bit, it's $3.3\times$ lower compared to the baseline and $5\times$ lower in our design.
Overall, we reduce area by $50\%$ for 3-bit and $42.2\%$ for 4-bit, with corresponding power savings of $33.3\%$.
It is important to note that the values for the Flash ADC include the resistor ladder, comparators and the encoder.

\begin{figure}[!t]
\centering
\includegraphics[width=\columnwidth]{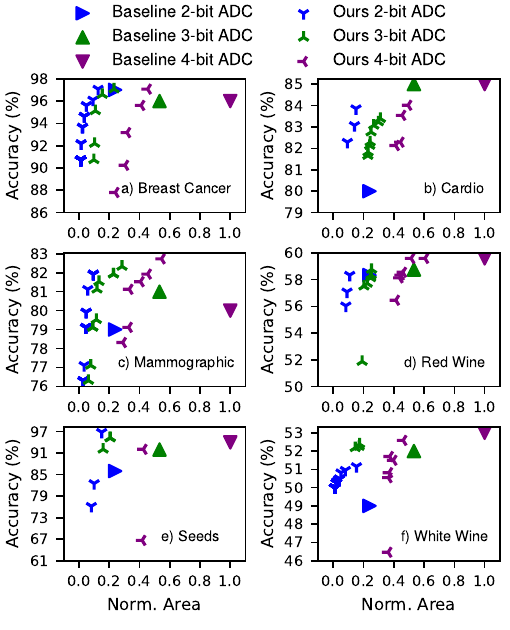}
\vspace{-6ex}
\caption{Pareto space of Accuracy vs normalized Area of 2, 3 and 4-bit ADCs.}
\vspace{-2ex}
\label{fig:pareto}
\end{figure}

\subsection{Partial ADCs Evaluation}\label{sec:comparison}

Next, we evaluate our proposed high-level pruning method in designing efficiently partial ADCs.
For our evaluation $2, 3 \textit{ and } 4$-bit ADCs considered in our analysis.
Figure~\ref{fig:pareto} illustrates how the suggested pruning process reduces the tailored ADC area over accuracy for \red{$2, 3, 4-\textit{bit}$} ADCs.
The area values are generated using our area model i.e. transistor count, whilst the accuracy refers to the classification accuracy of the respective ADCs' classifier.
The area values are normalized w.r.t. the area of the corresponding conventional Flash ADCs.
It should be noted that all prior works did not explore any ADC optimization and the conventional 4-bit Flash ADCs were used.
For the $2-bit$ ADC and up to 5\% accuracy degradation, the area gains in transistor count from \red{$2.4\times$} for \red{Seeds} and go up to \red{$24.2\times$} for \red{WhiteWine}.
For the $3-bit$ ADC and up to 5\% accuracy degradation, the area gains in transistor count from \red{$2.3\times$} for \red{Cardio} and go up to \red{$8.4\times$} for \red{Mammographic}.
For the $4-bit$ ADC and up to 5\% accuracy degradation, the area gains in transistor count range from \red{$2.4\times$} for \red{Cardio} and go up to \red{$3.5\times$} for \red{Mammographic}.
We should mention that $5\%$ accuracy loss is aligned with the accuracy requirements of our target applications.
Further, it should be highlighted that our approach not only explore efficiently the pareto space of the ADC w.r.t. the ADC area but also improves the overall accuracy of the classification system compared to the baseline accuracy.
As can be seen in \figurename~\ref{fig:pareto} our approach i.e., using partial ADCs, achieves higher classification accuracy than using the respective full binary ADC. 
As mentioned before, our approach by exploring the important ADC representations essentially targets to find the ideal ADC that follows the inputs distribution. 
Thus \figurename~\ref{fig:pareto} verifies that not only our approach reduce the ADCs area but further helps the quantization-aware training to converge to better solutions in terms of accuracy.

\begin{table}[]
\caption{Evaluation of the baseline printed MLPs.}
\vspace{-2ex}\label{fig:resultsMLPs}
\scalebox{0.7}{\begin{threeparttable}
\begin{tabular}{c|cc|ccc|c|c|c|}
\cline{2-9}
\rowcolor[HTML]{C0C0C0} 
\cellcolor[HTML]{FFFFFF}{\color[HTML]{FFFFFF} }                    & \multicolumn{2}{c|}{\cellcolor[HTML]{C0C0C0}{\color[HTML]{000000} Accuracy}} & \multicolumn{3}{c|}{\cellcolor[HTML]{C0C0C0}{\color[HTML]{000000} Area   (TC)²}}                                  & \cellcolor[HTML]{C0C0C0}{\color[HTML]{000000} }                                                                                  & \cellcolor[HTML]{C0C0C0}{\color[HTML]{000000} }                                                                                  & \cellcolor[HTML]{C0C0C0}{\color[HTML]{000000} }                                                                                  \\ \cline{2-6}
\rowcolor[HTML]{E7E7E7} 
\multirow{-2}{*}{\cellcolor[HTML]{FFFFFF}{\color[HTML]{FFFFFF} a}} & \multicolumn{1}{c|}{\cellcolor[HTML]{E7E7E7}Baseline¹}        & Pruned       & \multicolumn{1}{c|}{\cellcolor[HTML]{E7E7E7}Flash} & \multicolumn{1}{c|}{\cellcolor[HTML]{E7E7E7}Binary} & Pruned & \multirow{-2}{*}{\cellcolor[HTML]{C0C0C0}{\color[HTML]{000000} \begin{tabular}[c]{@{}c@{}}(3)\\Gains   (\%)\end{tabular}}} & \multirow{-2}{*}{\cellcolor[HTML]{C0C0C0}{\color[HTML]{000000} \begin{tabular}[c]{@{}c@{}}(4)\\ Gains   (\%)\end{tabular}}} & \multirow{-2}{*}{\cellcolor[HTML]{C0C0C0}{\color[HTML]{000000} \begin{tabular}[c]{@{}c@{}}(5)\\Gains   (\%)\end{tabular}}} \\ \hline
\rowcolor[HTML]{FFFFFF} 
\multicolumn{1}{|c|}{\cellcolor[HTML]{FFFFFF}2bit}                 & \multicolumn{1}{c|}{\cellcolor[HTML]{FFFFFF}73}               & 78.2         & \multicolumn{1}{c|}{\cellcolor[HTML]{FFFFFF}423}   & \multicolumn{1}{c|}{\cellcolor[HTML]{FFFFFF}235}    & 134    & 56                                                                                                                               & 57                                                                                                                               & 316                                                                                                                              \\ \hline
\rowcolor[HTML]{E7E7E7} 
\multicolumn{1}{|c|}{\cellcolor[HTML]{E7E7E7}3bit}                 & \multicolumn{1}{c|}{\cellcolor[HTML]{E7E7E7}77}               & 78.0         & \multicolumn{1}{c|}{\cellcolor[HTML]{E7E7E7}1138}  & \multicolumn{1}{c|}{\cellcolor[HTML]{E7E7E7}523}    & 249    & 46                                                                                                                               & 48                                                                                                                               & 458                                                                                                                              \\ \hline
\rowcolor[HTML]{FFFFFF} 
\multicolumn{1}{|c|}{\cellcolor[HTML]{FFFFFF}4bit}                 & \multicolumn{1}{c|}{\cellcolor[HTML]{FFFFFF}76}               & 78.0         & \multicolumn{1}{c|}{\cellcolor[HTML]{FFFFFF}2676}  & \multicolumn{1}{c|}{\cellcolor[HTML]{FFFFFF}981}    & 474    & 37                                                                                                                               & 48                                                                                                                               & 565                                                                                                                              \\ \hline
\end{tabular}

\begin{tablenotes}\normalsize
\vspace{0.5ex}
\item[] 
$^1$Accuracy for both Flash and Binary Baseline ADCs. 
$^2$Area in Transistor Count. 
$^3$Area gains by replacing Flash with Binary ADC. 
$^5$Area gains by replacing Binary with Binary Pruned ADCs. 
$^4$Area gains by replacing Flash with Binary Pruned ADCs. 
\vspace{-3.5ex}
\end{tablenotes}
\end{threeparttable}
}
\vspace{-2ex}    
\end{table}

Moreover, \ref{fig:resultsMLPs} presents the pareto points i.e., pruned Binary ADC designs, that has the higher accuracy for all datasets of \figurename~\ref{fig:pareto} for $2,3,4$-bits.
Table\ref{fig:resultsMLPs} conclude the effectiveness of our approach by presenting the accuracy has increased while the transistor count has decreased by moving from flash to full binary ADC and then finally to pruned binary ADC.
For $2,3,4$-bits ADC the accuracy increased on average $3\%$.
Further for $2,3,4$-bit resolution the ADC gains by shifting to Flash to Fully Binary ADC are on average $46\%$.
Our in-train pruning ADC approach achieved a further reduction of $51\%$ on average by replacing Fully Binary ADC with our partial Binary ADCs.
Finally, by reiterating that all previous works used a 4-bit Flash ADCs for their designs our approach compared to the corresponding Flash ADCs achieved a transistor count reduction of $446\%$ on average.

\section{Conclusion}\label{sec:conclusion}

Flexible Electronics (FE) represent promising avenues for low-cost production of electronic devices, particularly in applications like smart sensors, and wearables.
However, the integration of Analog to Digital Converters (ADCs) in these systems poses considerable challenges, particularly with respect to area and power consumption. 
We propose two different approaches to achieve optimal designs. 
This work addresses these issues by introducing a novel Binary Search ADC design that significantly reduces area overhead compared to existing solutions, achieving up to a $5.4\times$ reduction compared to Flash ADCs, and $2\times$ compared with state-of-the-art Binary design.  
Furthermore, our innovative in-training ADC optimization approach effectively minimizes the transistor count of ADCs by $5\times$, with negligible impact on accuracy. 
These advancements represent a critical step toward realizing efficient and compact flexible devices, thereby enhancing the practical application and performance of FE in targeted domains.

{\small
\section*{Acknowledgment}
This work is partially supported by the European Research Council (ERC), and co-funded by the H.F.R.I call “Basic research Financing (Horizontal support of all Sciences)” under the National Recovery and Resilience Plan “Greece 2.0” (H.F.R.I. Project Number: 17048).
}
\bibliographystyle{ACM-Reference-Format}
\bibliography{references.bib}

\end{document}